# Hopping mechanisms, photoluminescence studies toward highly efficient UV-responsive $Pr_2MgTiO_6$ photocatalyst

Moumin Rudra*, T. K. Bhowmik, H. S. Tripathi, R. A. Kumar, R. Sutradhar, and T. P. Sinha

*Department of Physics, Bose Institute, 93/1, A. P. C. Road, Kolkata, India 700009.*

*Corresponding author:

Address: *Department of Physics, Bose Institute, 93/1, A. P. C. Road, Kolkata, India 700009.*

Email: iammoumin@gmail.com

Phone: +91-33-23031194, +91-9477398318 (Mobile).

***ABSTRACT***

Titanium based-perovskites have got surprising attraction due to their environment-friendly photocatalysts behavior. The Rietveld refinement of X-ray diffraction (XRD) pattern authenticates the orthorhombic *Pnma* structure of $Pr_2MgTiO_6$ (PMT). The $Pr^{3+}$-$Ti^{4+}$ intervalence charge transfer process plays an important role inside PMT. The dc conductivity shows a change in the hopping mechanism from nearest-neighbor hopping to a variable range hopping due to the activation energy reduction. This change in the hopping mechanism is well supported by both the relaxation time and impedance data in PMT. The constant phase element (CPE) model is utilized to fit impedance plots. The photocatalytic activity of PMT was performed under UV irradiation with the help of degradation kinetics of organic pollutants. Also, the photocatalytic efficiency of recycled and reused PMT finds 91.2% after $5^{th}$ cyclic runs. The calculated band-gap is found to be 3.58 eV.

## 1. INTRODUCTION

Today, our main concern about global pollution, as day by day global industrialization developed rapidly. Organic pollutants are unable to remove completely using our traditional water treatment technology [1,2]. Only photocatalysts can overcome this problem, as it uses solar energy directly and it is also environment friendly. The titanium (Ti) based oxides are the ideal photocatalysts and widely used as photodegradation of organic pollutants [3-7], hydrogen production by water splitting [8], gas sensors [9], and solar cell [10]. Now a day, with increasing global civilization, we are searching for a material which has not only photodegradation power but also has different interesting properties. In recent years, Shawky et. al. reported about the photocatalytic performance of Ag/$LaTiO_3$ can retain 96% after fifth cycle [11]. This reusability and effective photocatalytic performance suggests $RETiO_3$ [RE = rare-earth lanthanides] is an ideal host material [12 - 15].

Previous reports on the $RETiO_3$ lanthanide-based perovskites show low-temperature Jahn–Teller cooperative distortions, which are characteristics of localized electrons, which can affect both the valences of RE and Ti cations results in improvement of transport and optical behaviors of material [16]. Also in $RETiO_3$, the $Ti^{4+}$ $3d^0$ orbital plays an important role to constitute the conduction band of the material [17]. Concerning other perovskites [18], an attractive structure makes $La_2MgTiO_6$ an ideal host material [19]. However, La is now being replaced by Pr as the electronic configuration of Pr matches with that of La and this replacement opens the $Pr^{3+}$–$Ti^{4+}$ intervalence charge transfer (IVCT) process in the material will influence the electron-hole population and further improve the transport properties of the material. Also, $Pr^{3+}$ has a unique $4f^2$ electronic configuration, which affects interesting properties in different applied areas like lighting devices [20], up-conversion areas [21], in-vivo targeted bio-imaging [22], etc. Also, another literature on $Pr_2NiTiO_6$ [23] suggests this type of material is also exhibits better electrical transport properties. In this work, the correlations between the $Pr^{3+}$and $Ti^{4+}$based structural factors, optical, and transport properties are discussed.

The bulks and grain-boundaries (*Gb*s) are the two main constituents, which make the material microstructures. In this study, we have used impedance spectroscopy to investigate the electrical transport characteristics associated with the bulks and *Gb*s contributions in $Pr_2MgTiO_6$ (PMT). Various hopping models (mobility activated charge carriers) have been presented to report the hopping mechanisms for DPO systems. Nearest-neighbour hopping is a type of hopping mechanism in which single-valued activation energy is involved, which contributes the carriers to hop on the adjacent neighbor site. Another type of hopping mechanism is Mott's variable range hopping (MVRH) mechanism, which occurs at low temperatures due to lowering activation energy with the decrease in temperature [24]. No attempt has been made to study the electrical transport properties as well as the optical and photochemical properties of the perovskite oxide PMT.

## 2. EXPERIMENTAL

### 2.1. MATERIALS AND METHODS

PMT in powder form was synthesized by a standard solid-state synthesis technique at 1573 K in the air for 14 hrs. using $Pr_2O_3$ (Sigma-Aldrich, 99.9%), MgO (Merck, 99%), and $TiO_2$ (Merck, 99.99%) as the ingredient materials. The polyvinyl alcohol (PVA) (5%) was added with the calcined powders and the mixture was pressed inside the cylindrical pelletizer. The pressed pellets were sintered at 1623 K to obtain dense PMT pellets with thickness, $d \approx 1$ mm & diameter ≈ 8 mm. The crystallographic structure of PMT was studied using a powder X-ray diffractometer (Rigaku Miniflex II) [20 kV, 15 mA, Cu-$K_\alpha$ ($\lambda$ = 1.5406 Å) radiation] in the 2θ range of 10° – 80° by the scanning rate of 0.02° per step at room temperature. A dual-position graphite diffracted beam monochromator has been used with a scintillation counter detector. The collected XRD data were refined with the help of the FULLPROF program using the Rietveld method [26]. During the Rietveld refinement, the pseudo-Voigt functions were used to fit peak shapes and the background was fitted using a 6-coefficient polynomial function. The microstructural image of the sintered pellet was captured using a scanning electron microscope (SEM) (FEI Quanta 200). JASCO FP 8500 was used to measured photoluminescence spectra. Shimadzu UV 2401 was used to collect the UV-vis spectrum. The Fourier transform infrared (FTIR) spectrum of PNT was recorded in transmittance mode within the range from 390 $cm^{-1}$ to 2000 $cm^{-1}$ with Perkin Elmer Spectrum 1000. The impedance spectroscopy study of the sintered pellet was measured in the temperature range from 300 K to 700 K (frequency range ≈ 45 Hz to 5 MHz), using an LCR meter (HIOKI – 3532, Japan). Before starting the experiment, both bases of the measuring pellet were cleaned and the contacts were prepared from the thin silver paste. The real ($\varepsilon'$) and imaginary ($\varepsilon''$) parts of the complex dielectric constant $\varepsilon^*$ (= $\varepsilon' + j\varepsilon''$, where $\varepsilon' = C_s/C_0$ and $\varepsilon'' = G/\omega C_0$) were obtained from the capacitance ($C_s$) and the conductance ($G$), whereas the real (Z') and imaginary (Z'') parts of the complex impedance Z* (=$Z' - jZ''$, where $Z' = Z\cos\varphi$ and $Z'' = Z\sin\varphi$) were obtained from the impedance (Z) and the phase angle ($\varphi$) [where ω is the angular frequency ($\omega = 2\pi\nu$), ν is the measured frequency and $j = \sqrt{(-1)}$. $C_0 = \varepsilon_0 A/d$ is the parallel plate capacitance of empty cell, A is the base area and $d$ is the thickness of pellet]. The ac electrical conductivity $\sigma$ (= $Gd/A$) was measured from the $G$. The photocatalytic activity was measured through the degradation of organic pollutant Rhodamine B (Rh-B) using a UV lamp. Before the UV lamp is on, 0.1 mg of sample was added with 100 ml of Rh-B (5 mg/L) solution. The solution is mixed in a dark cool bath to attain absorption-desorption equilibrium for 2 hr. In every 10 min interval, a Milex syringe (filter of 0.22 μm) was used to take 3 ml of solution from the solution bath for UV–Vis spectroscopy.

### 2.2. COMPUTATIONAL METHODS

The electronic structure of PMT has been calculated using the density functional theory (DFT) as implemented in the WIEN2k [25] package. The full potential linearized augmented plane wave (FP-LAPW) method is used to solve the Kohn-Sham equations. Firstly, the generalized gradient approximation (GGA) is taken as an exchange and correlation function to optimize the crystal structure. But it does not create any band-gap energy in the electronic structure because of the presence of the strongly correlated f-states of the Pr atom and d-states of the Ti atom. So, GGA along with the Coulomb repulsion, U (GGA+U) is finally adopted to optimize the structure and further calculations. However, the supercell has been formed with a size of 2*1*1 of lattice vectors, where one-unit cell consists of Mg atom and other Ti. The muffin-tin radii (RMT) for Pr, Mg, Ti, O are 2.48, 1.79, 1.88, and 1.70 a.u respectively. The energy cut off (RMT*$k_{max}$) is set to -8 Ry to core states from the valance states. The k-space integration has been carried out with 500 k-points over the whole Brillouin zone. The Hubbard parameter ($U_{eff} = U - J$) is taken to 6 eV for Pr-4*f* states and 3 eV for Ti-3*d* states, where the *J* is set to zero. To achieve the self-consistency solutions, the energy and charge convergence criteria are set to $10^{-4}$Ry and $10^{-3}$ e. After optimizing the crystal structure, the band structure and density of states (DoS) have been calculated for PMT double perovskite.

## 3. RESULTS AND DISCUSSION

### 3.1. X-RAY DIFFRACTION ANALYSIS:

The refined XRD pattern of the calcined powder of PMT at RT is shown in Fig.1 (a). The good quality of the refinement is found as the difference between the observed and the calculated pattern was quite small. The lower symmetric orthorhombic *Pnma* (#62) structure of the PMT was authenticated by the Rietveld refinement, which corresponding to the $a^+b^-b^-$ tilt system. The existence of superlattice reflection peaks (011), (210), (230/212), and (311) indicate the disordered orthorhombic structure. The refined parameters were found to be $R_{exp}$ = 6.77, $R_p$ = 5.39, $R_{wp}$ = 7.06 and $\chi^2$ = 1.09. Also, the lattice parameters, which are refined, are found to be $a$ = 5.5710 Å, $b$ = 7.7893 Å, and $c$ = 5.4889 Å. The inset of Fig. 1 shows the PMT unit cell with the sharing of atoms at the Wyckoff positions 4*c* (*x*,¼,*z*) for $Pr^{3+}$ atoms, 4*b* (½,0,0) for $Mg^{2+}$ and $Ti^{4+}$ atoms, 4*c* (*x*,¼,*z*) and 8*d* (*x*,*y*,*z*) for $O^{2-}$ atoms. Each B-site cation ($Mg^{2+}$ or $Ti^{4+}$) is bonded with six $O^{2-}$ atoms to constitute $MgO_6$ and $TiO_6$ octahedra, respectively. The crystallographic refined parameters are tabulated in Table 1. Fig. 1 (b) displays the SEM image of the PMT pellet. The SEM image of PMT discloses the bulks of various sizes and shapes. The domain structure of the PMT with the "lamellae" character is observed. The size of the bulk particles (i.e., grains) varies from 1.25 μm to 4.7 μm.

### 3.2. LUMINESCENCE STUDIES:

Fig. 2(a) and 2(b) show the photoluminescence excitation (PLE) and emission (PL) spectra of PMT at 300 K (RT), respectively. The PLE spectra are monitored at 200 nm and 613 nm. From Fig. 2(a), it is seen that the PLE spectra consist of two broad bands are located around 277 nm (A band) and 367 nm (B band), are assigned to host absorption of the PMT lattice [26–28] and $Pr^{3+}$–$Ti^{4+}$ IVCT absorption [29–31], respectively. The PL spectra of PMT show a series of emission peaks in the 400–575 nm range when it is excited with 200 nm, 277 nm (A band), and 367 nm (B band). From the PL spectrum, near Band Edge emission (NBE) was calculated and the band gap energy is found to be 4.48 eV from Tauc plot [Inset of Fig. 2(a)]. From Fig. 2(b), the series of emission peaks consists of a strong band in the range 400–450 nm, a peak at 470 nm, and a weak band in the range 485–550 nm, which corresponding to $^3H_4 \rightarrow {}^3P_J$ (J = 2, 1, 0), $^1I_6$ (~ 400, 415, 422, 438 nm), $^3P_0 \rightarrow {}^3H_4$ and $^3P_1$, $^3P_0 \rightarrow {}^3H_5$ (~ 485, 515 nm) transitions of $Pr^{3+}$ atoms [32, 33], respectively.

Fig. 2(c) shows the schematic energy level diagram of the PMT. This diagram helps to clearly understand the excitation and recombination processes in PMT. The electrons go to the $^3P_2$ state of $Pr^{3+}$ from its ground state $^3H_4$ upon 400 nm excitation. Then the electrons relax rapidly to $^3P_0$ and $^1D_2$ lower states via IVCT, i.e., the electrons are trapped by $Ti^{4+}$ orbital before they recombine with the holes ($h^+$) on $Pr^{3+}$ ground state. During the IVCT process, the electrons are excited to $Ti^{4+}$ orbital from the ground state ($^3H_4$) of $Pr^{3+}$ and created [$Ti^{4+}$+$e^-$] state [34]. The excited electrons are relaxing to the lowest potential energy of [$Ti^{4+}$+$e^-$] state through a fast non-radiative process. Lastly, the electrons are back to the $Pr^{3+}$ ground state via the excited ($^3P_0$, $^1D_2$) states of $Pr^{3+}$. In summary, the electron trapping and it recombines with the holes can be written as follows:

$$\text{PMT} + h\nu \rightarrow e^- + h^+ \quad (1)$$

$$Pr^{3+} + h^+ \rightarrow Pr^{4+} \quad (2)$$

$$Ti^{4+} + e^- \rightarrow Ti^{3+} \rightarrow Ti^{4+} + e^- \quad (3)$$

$$Pr^{4+} + e^- \rightarrow (Pr^{3+})^* \quad (4)$$

$$(Pr^{3+})^* \rightarrow Pr^{3+} + \text{PL} \quad (5)$$

where $(Pr^{3+})^*$ represents excited states of $Pr^{3+}$ atom, (Eq. 1) the light falls on the PMT creates free electron ($e^-$) and holes ($h^+$) in the conduction band (*cb*) and valence band (*vb*), respectively. (Eq. 2, 3) The electrons are trapped by the $Ti^{4+}$ and creates [$Ti^{4+}$ + $e^-$] of $Ti^{3+}$ state, on the other side holes ($h^+$) are trapped in $Pr^{3+}$ and creates $Pr^{4+}$ state. (Eq. 3, 4) The $Ti^{3+}$ or [$Ti^{4+}$ + $e^-$] state releases the *cb* electron, later which are recombines with the $Pr^{4+}$ and creates the excited $(Pr^{3+})^*$ state [$^3P_J$, $^1D_2$]. (Eq. 5) Finally, the excited $(Pr^{3+})^*$ state relaxes to the ground state [$^3H_4$] of $Pr^{3+}$, which results in the weak and strong bands in the emission spectra.

The absorption energy of IVCT can be estimated by the following equation [35]:

$$\text{IVCT (cm}^{-1}) = 58800 - 49800[\chi(Ti^{4+})/d(Pr^{3+}\text{–}Ti^{4+})]$$

where $\chi(Ti^{4+})$ is the optical electro-negativity (2.05) of $Ti^{4+}$ atom; $d(Pr^{3+}\text{–}Ti^{4+})$ is the smallest bond length (3.21646 Å) of $Pr^{3+}$–$Ti^{4+}$ in the PMT lattice, which obtained from Rietveld refinement data. Thus, the absorption energy of IVCT is calculated to be 27060 $cm^{-1}$ (~ 3.3550 eV ≈ 369 nm), which is approximately similar to the observed data 27248 $cm^{-1}$ (~ 3.3783 eV ≈ 367 nm) obtained from the PLE spectra. Also, we have calculated the band-gap ($E_g$) of the PMT using the Tauc plot (direct allowed transition) (Inset of Fig. 2(d)) of UV-vis spectrum (Fig. 2(d)) and the value of $E_g$ is found to be 4.49 eV (~ 277 nm), which coincides with the host absorption value (A band) in the PLE spectra.

### 3.3. FOURIER TRANSFORM INFRARED (FTIR) ANALYSIS:

Fig. 3 shows the FTIR spectrum of PMT. All the peaks and bands are the characteristic of the material except the hump at 1600–2000 $cm^{-1}$, which is due to the presence of absorbed moisture in KBr [$KBr.(H_2O)_m$]. The first two peaks at 397 $cm^{-1}$ and 419 $cm^{-1}$ may be attributed to the asymmetric bending modes of $MgO_6$ and $TiO_6$ octahedra respectively. Whereas the peaks at 431 $cm^{-1}$ and 457 $cm^{-1}$ corresponds to symmetric bending modes of $MgO_6$ and $TiO_6$ octahedra respectively. In Fig.3, the inset displays peaks at 570 $cm^{-1}$ and 605 $cm^{-1}$, which correspond to asymmetric stretching of $MgO_6$ and $TiO_6$ octahedra respectively, whereas the two peaks at 930 $cm^{-1}$and 975 $cm^{-1}$ may be attributed to the symmetric stretching modes of $MgO_6$ and $TiO_6$ octahedra respectively. The small intensity peaks found in the range 1020–1520 $cm^{-1}$, likely correspond to the overtones of the fundamental vibrations in PMT. In FTIR spectra, we generally get two peaks for the two types of octahedra in PMT, which can be explained by the following equation, where the wavenumber of transmittance (1/λ) can be expressed as

$$\frac{1}{\lambda} = \frac{\nu}{c} = \frac{1}{2\pi c}\sqrt{\frac{k}{\mu}}$$

where $k$ is the molecular force constant, $c$ is the speed of light in the vacuum (i.e., 299792458 m/s) and $\mu$ is the reduced mass of the Ti/Mg–O system. The reduced masses for Ti–O and Mg–O bonds which come out to be 11.9911u and 9.648u respectively, where 1u = $1.660539040(20) \times 10^{-27}$ kg. As the reduced mass of Ti–O bonds are larger than Mg–O bonds, therefore the transmittance peaks appear in the lower wavenumber side for Ti–O bonds.

## 3.4. CONDUCTIVITY ANALYSIS:

The ac conductivity data was measured over the temperature range 300 (RT) – 700 K. The variations of ac conductivity with the frequency (ν) at various temperatures are shown in Fig. 4(a). In the measured frequency range, the ac conductivity spectra show two different regions, which are due to the effects of grain-boundaries (*Gb*) at low frequency (Region I) and that of bulks at high frequency (Region II) [36]. The dc conductivity value ($\sigma_{dc}$) can be estimated by extrapolating conductivity spectra at ν = 0. The whole ac conductivity shows a movement towards the high-frequency side with the increasing temperature. We have adopted the double power law to fit the conductivity spectra and highlight the contributions of *Gb*s and bulks [36]

$$\sigma_{ac} = \sigma_{dc} + A\omega^{k_1} + B\omega^{k_2} \quad (6)$$

where $A$, $B$ are the temperature-dependent constant, whereas $k_1$, $k_2$ are the material-dependent parameters. The fitted ac conductivity spectra for 700 K are shown in Fig. 4(a). The fitted parameter values are tabulated in Table 2.

Fig. 4(b) shows the temperature-dependent $\sigma_{dc}$for PMT, which shows a decrease in $\sigma_{dc}$ with increasing temperature suggesting the semiconducting nature of PMT. The values of $\sigma_{dc}$ are collected from the previous fitted values of $\sigma_{ac}$ and found approximately the same when matched with the observed data of $\sigma_{dc}$. In Fig. 4(c), the values of $\sigma_{dc}$ are fitted using the nearest-neighbor hopping (NNH) model, defined as,

$$\sigma_{dc} = \sigma_\alpha \exp\left(-\frac{E_a}{k_B T}\right) \qquad (7)$$

where$\sigma_\alpha$ is the pre-exponential constant, Boltzmann constant ($k_B$), the activation energy($E_a$), and the temperature ($T$). It is seen that the observed data of $\sigma_{dc}$ at the high-temperature region is well fitted by Eq. 7, whereas at 400 K, a change in its linear behavior is found, which suggests another dissimilar hopping mechanism in this low-temperature region (below 400 K). From the linear behavior in Fig. 4(c), the value of $E_a$ is obtained and found to be 0.33 eV, which suggesting polaronic conduction between the nearest neighboring atoms [37]. To investigate the type of hopping mechanism below 400 K, the value of $E_a$ at each temperature is calculated with the equation $E_a = - d(\ln\sigma)/d(1/k_B T)$ [38] and the inset of Fig. 4(c) shows the temperature variation of $E_a$. It is found that the value of $E_a$ decreases to 0.275 eV from its initial value of 0.33 eV with the decrease in the temperature below 400 K, which suggests the range of hopping distance is large than the nearest-neighbor distance due to decreasing activation energy. This decreasing trend of activation energy suggests the disordered states of PMT due to charge imbalance or IVCT between $Pr^{3+}$ and $Ti^{4+}$ in sample.

In this point of view, Mott's variable-range hopping (MVRH) is adopted to fit the values of $\sigma_{dc}$ in the lower temperature region due to disordered states of PMT are shown in Fig. 4(d). [39, 40]

$$\sigma_{dc} = \sigma_\beta \exp\left[-\left(\frac{T_0}{T}\right)^{\frac{1}{4}}\right] \tag{8}$$

where$\sigma_\beta$ is the pre-exponential constant and the characteristic temperature ($T_0$). The value of $T_0$ can also be defined as follows:

$$T_0 = \frac{18}{k_B N(E_F)\xi^3} \tag{9}$$

where the $N(E_F)$ is the density of states at the Fermi energy and $\xi$ be the localization length. The fitted parameter $T_0$ is found to be 669348164 K. The hopping energy $E_h$ is given by $\frac{1}{4}k_B T^{\frac{3}{4}} T_0^{\frac{1}{4}}$[41]. It is found that the $E_h$ increases to 0.31 eV from 0.25 eV [inset of Fig. 4(d)].By adopting a projected value (2.2 Å) of ξ [41], the $N$ ($E_F$) value is measured to be $2.93\times10^{25}$ $eV^{-1}m^{-3}$, suggesting a higher concentration of charge carriers as reported in a similar type of materials [41]. The hopping distance $R_h$is also calculated by $\frac{3}{8}\xi\left(\frac{T_0}{T}\right)^{\frac{1}{4}}$[42] and it is decreasing to 2.967 nm from 3.188 nm with the increasing temperature. From these results, we can conclude that the $\sigma_{dc}$ points towards a transition from NNH to MVRH around 400 K with the decrease in $E_a$.

### 3.5. DIELECTRIC CONSTANT:

Fig. 5(a) and 5(b) shows the frequency variation of dielectric constant (ε′) and dielectric tangent loss (tanδ) in the temperature range of 300–700 K. One dispersion region is observed in Fig. 5(a) and also one prominent relaxation peak in tanδ is present in the measured frequency range, which indicates the contribution of bulks in the material. On the low-frequency side, the tail of another relaxation process is observed due to *Gb*s contribution, which is not observed at RT data (300 K). This suggests only bulks contribution dominates the dielectric data at 300 K in the measured frequency range, whereas above RT, the *Gb*s contribution is also present along with the bulk contribution. This ε′ dispersion region and the corresponding tanδ peak move towards the high-frequency side with the increase in temperature, which suggests a thermally activated process. The charge carriers' mobility and the temperature strongly dominate the tanδ peaks. So, the mobility is increased with the increase in temperature, which results in the movement of tanδ peak towards the high-frequency side. The values of tanδ peaks for bulks is around 0.25 - 0.5, which is observed above $10^3$ Hz and that for *Gb*s is greater than 0.5 at the low-frequency side. A similar value of tanδ peak for bulks is reported in various literatures [43, 44].

Fig. 5(c) shows the variation of dielectric constant (ε′) with the temperature of PMT at various frequencies. It is observed that the values of ε′ increases with the temperature. As the number of charge carriers increases with the increasing temperature, which helps increase in polarization, results increase in ε′ value also. The ε′ values increases stair-like with the temperature (Inset of Fig. 5(c)), which is due to the presence of two microstructural effects (i.e., bulks and *Gb*s) in the material, it is quite similar in the materials with a high dielectric constant [45-50]. Here, *Gb*s have higher capacitance than that of bulks. This *Gb*s creates a highly capacitive layer to traps more charge carriers, which results in a high ε′ above RT at the low-frequency side [51, 52].

The variation of relaxation times (τ) with temperature are plotted in Fig. 5(d) by utilizing the NNH model.

$$\tau = \tau_o \exp\left(\frac{E_a}{k_B T}\right) \quad (10)$$

where $\tau_o$ is the pre-exponential factor. The values of τ are collected from the tanδ peak positions in Fig. 5(b). These peak positions are collected only for bulks' contribution as the tanδ peaks for *Gb*s are not observed in the measured frequency range. From Fig. 5(d), it is observed that the points (τ values) fit well with the straight line, and the value of $E_a$ is found to be 0.25 eV for this region. On the low-temperature side, the τ values deviate the straight-line fit suggesting the presence of another mechanism below 400 K. In this low-temperature region, we have employed the MVRH model (Inset of Fig. 5(d)) using the equation as given below:

$$\tau = \tau_\alpha \exp(T/T_0)^{1/4} \tag{11}$$

where $\tau_\alpha$ is also a pre-exponential constant. We have seen that a good agreement between the τ values and MVRH is obtained. This result supports the outcomes from dc conductivity data as discussed in Sec.3.4.

### 3.6. IMPEDANCE ANALYSIS:

The complex impedance plane plots (i.e., Z-plot) for PMT at various temperatures are shown in Fig. 6(a) & 6(b). The Z-plot shows one depressing semicircular arc with the tails of another semicircular arc suggesting the presence of two different microstructural effects in PMT. From Fig. 6(a) & 6(b), it is seen that the diameter of semicircular arcs are decreasing with the temperature, which throws light on the semiconducting behaviour of sample. As discussed in Sec. 3.5, the *Gb*s is less conducting than bulks, so that the contribution of bulks always lies at relatively high frequencies in the material. The decrease in diameter of these semicircular arcs with the increase in temperature also suggests the thermally activated process [52]. The depressed semicircular arcs with the center below X-axis suggest the non-ideal capacitance in PMT. So, we have employed the constant phase element (CPE) model to fit these Z-plots. The non-ideal capacitance of the CPE model can be assumed as $C_{CPE} = Q^{1/n}R^{(1-n)/n}$, where $n$ ($0 \leq n \leq 1$) is a constant that depends on the non-ideal behavior of the capacitance. Here $Q$ and $R$ are the constant phase element and resistance. In Fig. 6(a) and 6(b), the observed data (color points) are fitted with the electrical circuit as shown in the inset of Fig. 6(a). The temperature variation of fitted parameter $n$ is shown in the inset of Fig. 6(b) and the $n$ values increase to 0.995 at 700 K from 0.96 at 300 K (RT). This suggests the bulk capacitance approaches to ideal behavior with an increase in temperature, which may be due to the removal of defects in the bulk interior. The fitted resistance $R$ decreases with an increase in temperature suggesting the semiconducting behavior of PMT.

To study the hopping mechanism from impedance data, we have also plotted the fitted parameter $R$ using the NNH model (Eq. 12) as shown in Fig.6(c).

$$R = R_{\alpha} e^{\left(E_a/k_B T\right)} \tag{12}$$

where $R_{\alpha}$ is a pre-exponential constant. The observed ($R$) points also fit well with the straight line above 400 K with a corresponding $E_a$ value is 0.235 eV for the bulks. We have found small differences in the activation energies from dc conductivity, relaxation time, and impedance data. As the hopping energy of the charge carriers are responsible for relaxation and impedance, whereas the disorder and binding energy of polarons along with the hopping energy are involved in the hopping mechanism [53-55]. From these results, we can summarize that similar charge carriers are responsible for the conduction, relaxation, and impedance process. The deviation of points ($R$-values) from the straight line suggests the presence of another hopping mechanism below 400 K in PMT. In this region, we have also employed the MVRH model (Fig. 6(d)) using the equation given below:

$$\ln(R/R_0) = (T/T_0)^{1/4} \tag{13}$$

where $R_0$ is a constant value of $R$. The MVRH model fits well with the fitted $R$-values, which also supports the outcomes from dc conductivity and relation time data as discussed Sec. 3.4 &3.5.

### 3.6. PHOTOCATALYTIC ACTIVITY:

To study the photocatalytic activity of PMT, the photocatalytic degradation of the organic dye rhodamine B (Rh-B) was performed as per literature [56]. Fig. 7(a) shows the changes in absorbance peak around 550 nm with time in presence of UV-light for a maximum of 75 min. The color of the Rh-B solution changes with time and it became colorless after 75 min in presence of both UV-light irradiation and the PMT. Similar experiments were carried out in absence of UV-irradiation with the sample and the presence of UV-irradiation without the sample PMT. The percentage (%) of degradation was defined as $((C_0 - C)/C_0) \times 100\%$, where $C_0$ is the Rh-B solution's initial concentration and $C$ is the Rh-B solution's concentration during the photocatalytic reaction. From Fig. 7(b), it is found that the degradation (%) of the PMT was about 71%, whereas 0.9% and 3.3% degradations were found in absence of UV-light and PMT, respectively after 30 min. The kinetic behavior of Rh-B degradations are analyzed to understand the photocatalytic efficiency of the PMT sample. The reaction kinetics can be described in terms of first-order kinetics as following [57, 58]:

$$ln\frac{C}{C_0} = -K_a t \tag{14}$$

where the apparent rate constant ($K_a$) is the fundamental kinetic parameter for various photocatalysts. Fig. 7(c) shows $ln\frac{C}{C_0}$ against the time graph and the value of $K_a$ is found to be 0.05058 $min^{-1}$. For the industrial application, it is important to check the photocatalytic efficiency of the recycled and reused PMT by several times. The photocatalytic efficiency (% degradation) of PMT for 5 times reuses were 98, 96.8, 94.6, 93.9, and 91.2% after 75 min of reaction time, respectively (Fig. 7(d)). No leaching for Pr, Mg or Ti into water was found even after 5$^{th}$ cycle of experiment. There is no change in structural and PL behavior of PMT after each experiment.

In summary, the photocatalytic activity in the PMT can be described based on band alignment as follows:

$$\text{PMT} + h\nu \rightarrow h_{vb}^{+}.\text{PMT}.e_{cb}^{-} \quad (15)$$

$$h_{vb}^{+}.\text{PMT}.e_{cb}^{-} + O_2 \rightarrow h_{vb}^{+}.\text{PMT} + O_2^{*-} \quad (16)$$

$$h_{vb}^{+}.\text{PMT} + OH^{-} \rightarrow \text{PMT} + OH^{*} \quad (17)$$

$$O_2^{*-} + H_2O \rightarrow HO_2^{*} + OH^{-} \quad (18)$$

$$HO_2^{*} + H_2O \rightarrow OH^{*} + H_2O_2 \quad (19)$$

$$H_2O_2 \rightarrow 2\ OH^{*} \quad (20)$$

$$\text{Rh-B} + OH^{*} \rightarrow \text{degradation product} \quad (21)$$

(Eq. 15) The electrons in the *vb* can be excited to the *cb*, with the same number of holes in the *vb* upon UV irradiation. The electrons are trapped by $Ti^{4+}$ and the holes by $Pr^{3+}$ in the PMT. This charge carrier separation results increase in carrier lifetime due to less charge recombination rate. As a result, more hydroxyl $OH^{*}$ radicals are created by the reaction of trapped holes and electrons (Eq. 16, 17, 18, 19, 20). These strong oxidizing agents $OH^{*}$ radicals decompose the organic Rh-B dye (Eq. 21) results in colorless solutions. Therefore, $Pr^{3+}$ - $Ti^{4+}$ channel in PMT opens a new route to increase charge carriers' lifetime as well as less recombination rate.

## 3.7.DENSITY OF STATES AND BAND STRUCTURE:

The spin-polarized DoS of PMT unit cell is calculated using the GGA+U approximation is shown in Fig. 8(a). Also, for the knowledge about the individual contribution of Pr 4*f*, Pr 4*d*, Ti 3*d*, and O 2*p* states to the total DoS, the partial DoS of these states are shown in Fig. 8(b) – 8(d). The calculated band-gap (**E<sub>g</sub>**) is found to be 3.58 eV in Fig. 8(a) when the Fermi energy (**$E_F$**) is set at 0 eV. In PMT, the up-spin and down-spin channels are specified by ↑ and ↓, respectively. The spin-polarized band structures along with total DoS of PMT are shown in Fig. 9. The band structure on the right-hand side (red) is for the up-spin channel, whereas the band structure on the left-hand side (blue) is for the down-spin channel. It is found that the PMT shows a direct band-gap between conduction band minimum and valence band maximum at Γ-point for both up-spin and down-spin channels. The up-spin valence band is composed of Pr 4*f*/4*d* and O 2*p* states with a small contribution of Ti 3*d* state. On the downside, the O 2*p* state mainly contributes to the down-spin valence band with a smaller contribution of Ti 3*d* state. It is also observed that the O 2*p* state mainly contributes at the valence band maximum near 0 eV, whereas the contribution of Pr 4*f*/4*d* and Ti 3*d* states are nearly zero. The up-spin and down-spin conduction bands are consisting of Pr 4*f*/4*d*, Ti 3*d*, and O 2*p* states, but the Ti 3*d* ($t_{2g}$ and $e_g$) states mainly dominate the conduction band minimum with a smaller contribution of O 2*p* orbital. It is also observed that Ti 3*d* state splits into $t_{2g}$ and $e_g$ states by the crystal field formed by oxygen octahedra, with $t_{2g}$ state at lower and the $e_g$ at higher energy levels. So, the partial DoS of all states showing light on the bonding structure of PMT. The O 2*p* orbital mainly contributes at valence band maximum, whereas Ti 3*d* orbital at conduction band minimum, which suggesting the hybridization of Ti 3*d* and O 2*p* orbitals in PMT. The calculated band-gap (3.58 eV) is found to be smaller than the experimental band-gap (4.49 eV) in Sec. 3.2.

## 4. CONCLUSIONS:

In this work, the Rietveld refinement was performed on the observed X-ray diffraction (XRD) data, which authenticates the lower symmetric orthorhombic *Pnma* structure of solid-state synthesized PMT. The $Pr^{3+}$-$Ti^{4+}$ intervalence charge transfer process plays an important role inside PMT. The ac conductivity spectra are analyzed based on double power law. The dc conductivity shows a change in the hopping mechanism from NNH to MVRH due to the activation energy reduction. This change in the hopping mechanism is well supported by both the relaxation time and fitted *R* from Z-plots in PMT.The constant phase element (CPE) model is utilized to fit Z-plots. The bulk capacitance approaches ideal behavior with an increase in temperature, which is due to the removal of defects in the bulk interior. The same type of charge carrier is responsible for the conduction, relaxation, and impedance process. The photocatalytic activity of PMT was performed under UV irradiation with the help of degradation kinetics of rhodamine B. Also, the photocatalytic efficiency of recycled and reused PMT is found 91.2% after the 5$^{th}$ cyclic run, which is an important parameterfor the industrial application. The calculated band-gap of PMT is found to be 3.58 eV.

## 5. CONFLICT OF INTEREST

There is no conflict of interest exists.

## 6. ACKNOWLEDGMENT

MR (NET JRF, ID no. 522407) acknowledges the financial support provided by the University Grants Commission (UGC), India.

## 7. DATA AVAILABILITY STATEMENT

The data that support the findings of this study are available from the corresponding author upon reasonable request.

Figure

Manuscript number: **PHYSB-D-20-02735**

Title: **Hopping mechanisms, photoluminescence studies toward highly efficient UV-responsive $Pr_2MgTiO_6$ photocatalyst.**

**Figures:**

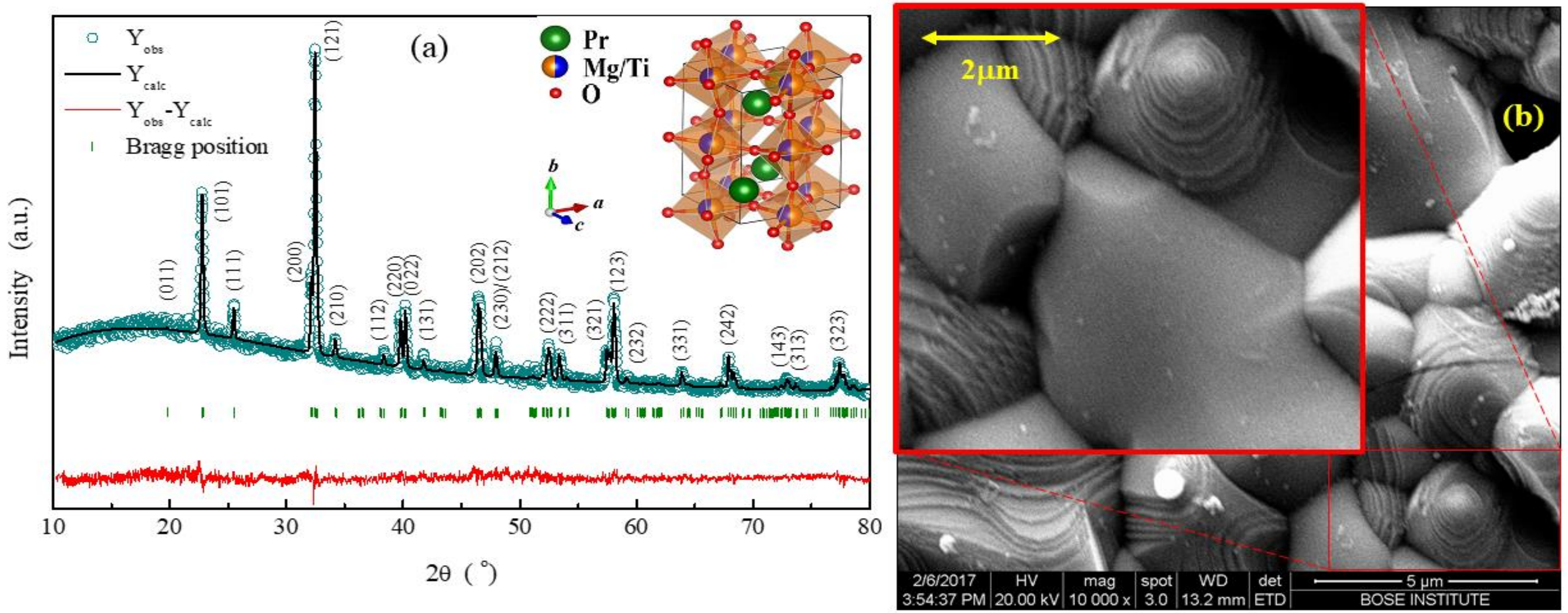

Fig. 1: (a) The X-ray diffraction pattern of PMT at RT, inset shows the unit cell of PMT. (b) The SEM micrograph of PMT.

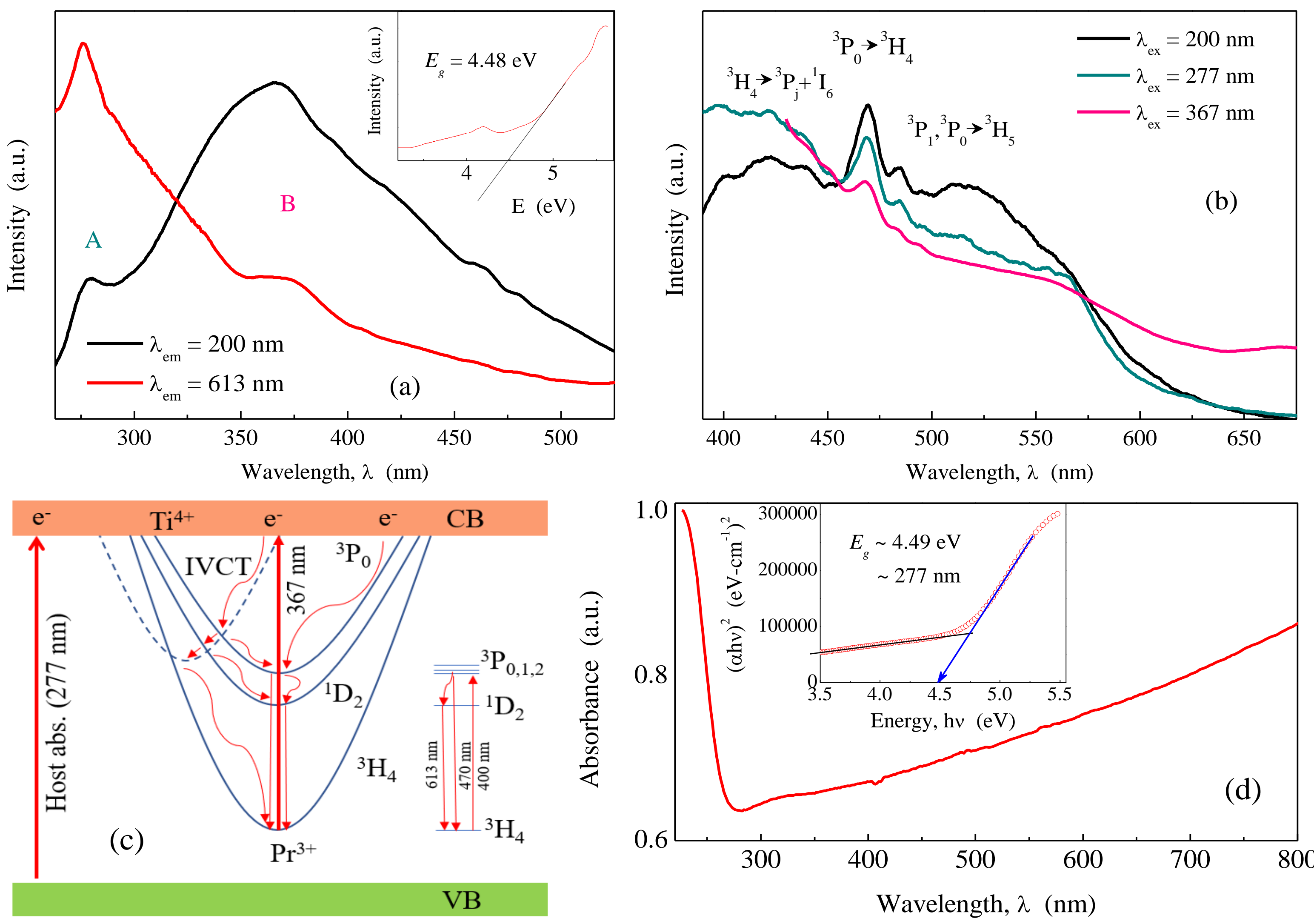

Fig. 2: The photoluminescence excitation (a) and emission (b) spectra of PMT at RT. (c) The schematic energy level diagram of the PMT. (d) The UV-vis absorbance spectrum of PMT and the inset shows the Tauc plot (direct allowed transition) calculated from the UV-vis spectrum.

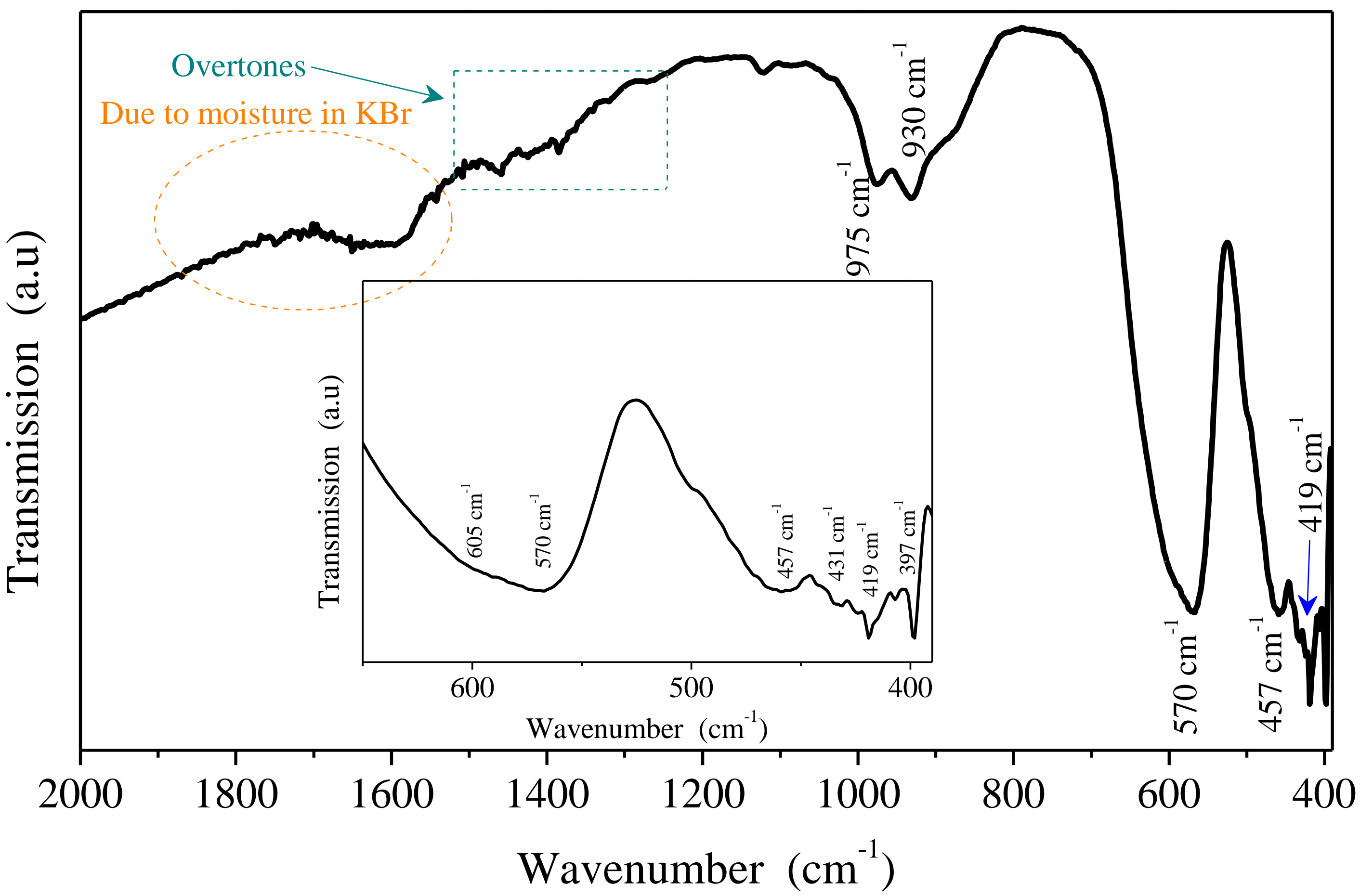

Fig. 3: The FTIR spectrum of PMT.

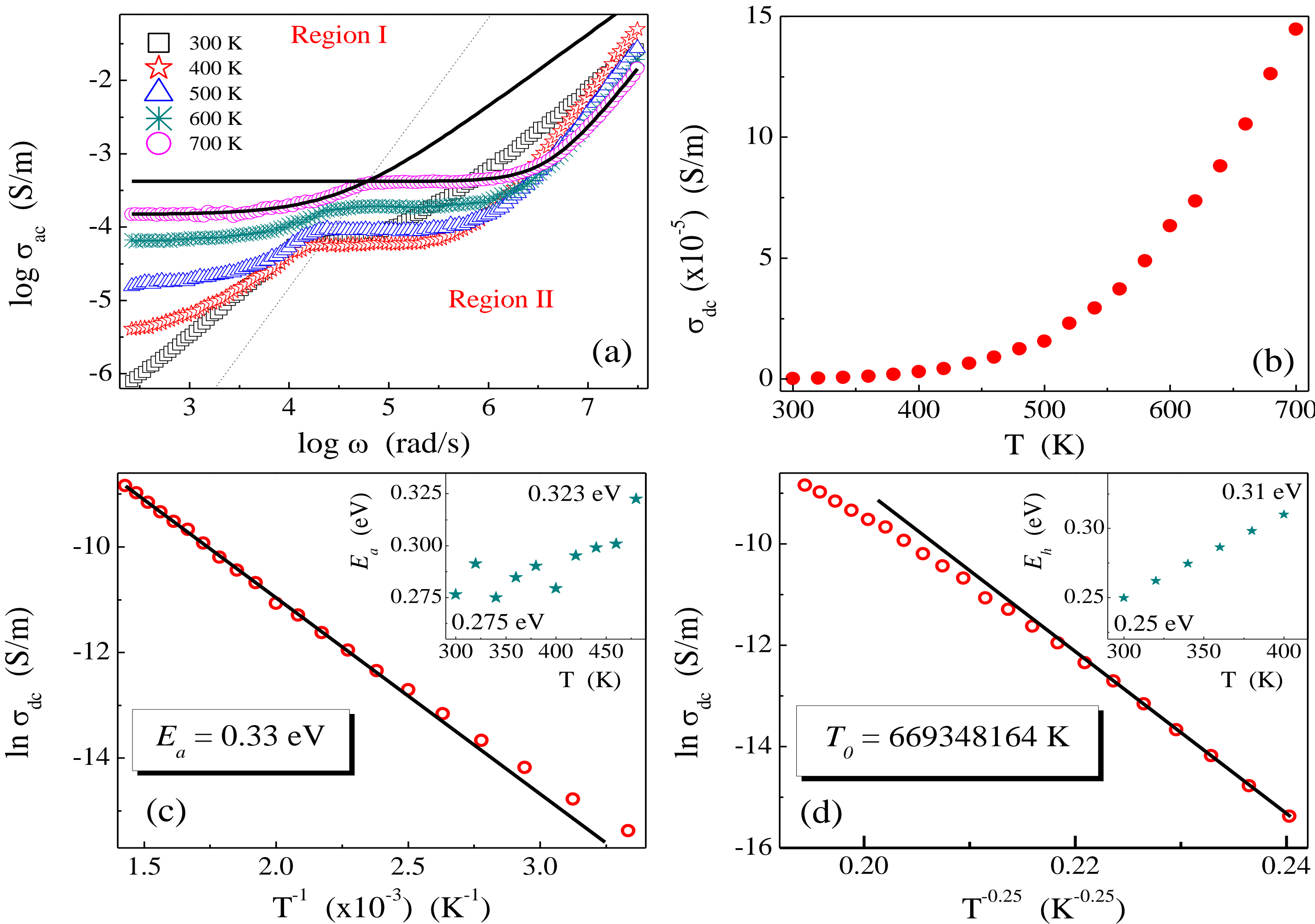

Fig. 4: (a) Frequency dependence of the ac conductivity at various temperatures. (b) The temperature variation of dc conductivity for PMT; The dc conductivity of PMT describes the conduction process using (c) NNH model, (d) MVRH model; The insets show the hopping energy (d) as a function of temperature; (c) Variation of activation energy with the temperature below 400 K.

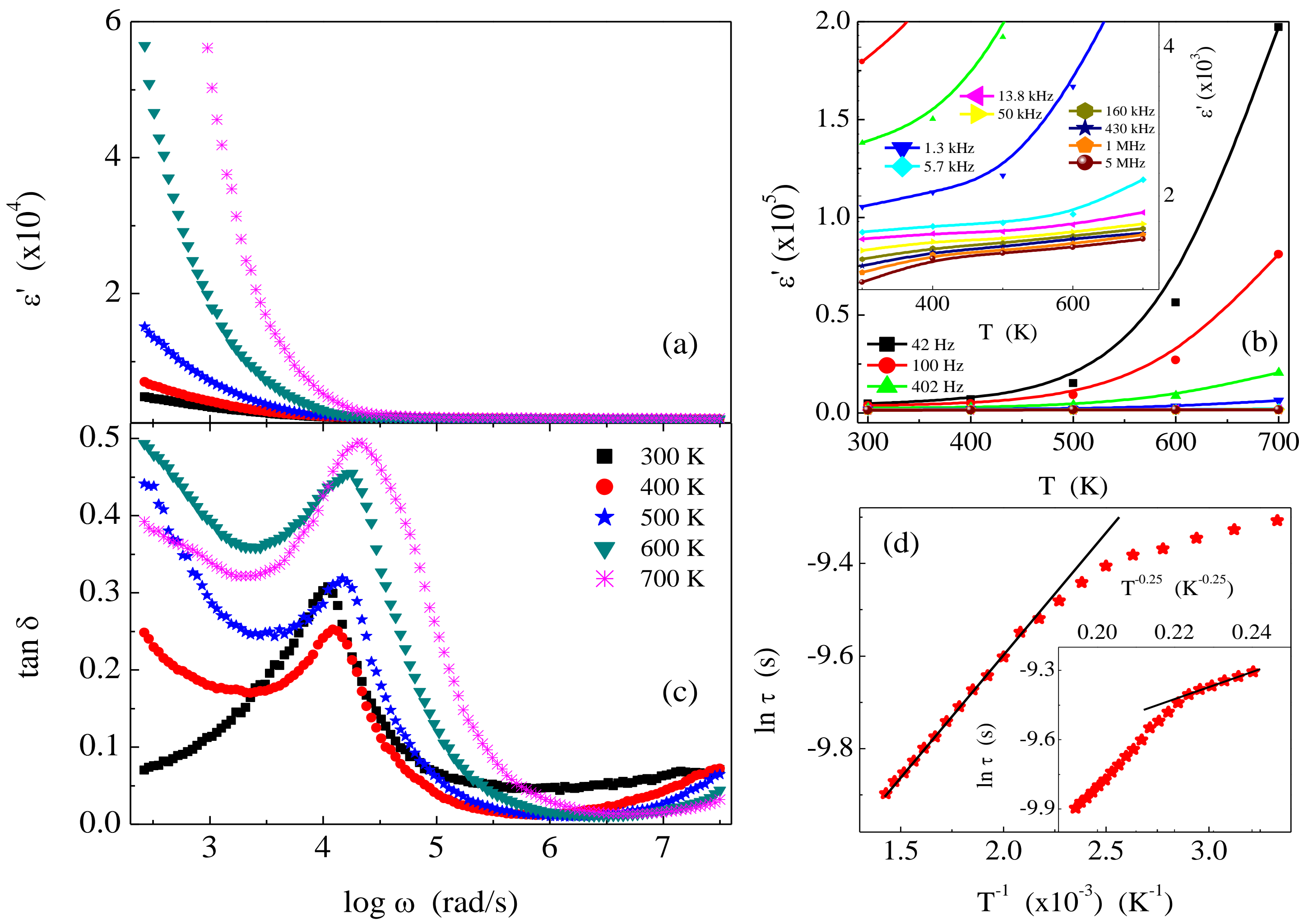

Fig. 5: Frequency dependence of ε′ (a) and tanδ (b) at various temperatures for PMT. Temperature dependence of ε′ (c) at various frequencies for PMT. Inset (c) shows the closed view of ε′ with temperature. Relaxation times τ of the charge carriers at bulks are plotted using (d) NNH model and the inset shows τ is plotted using (d) MVRH model.

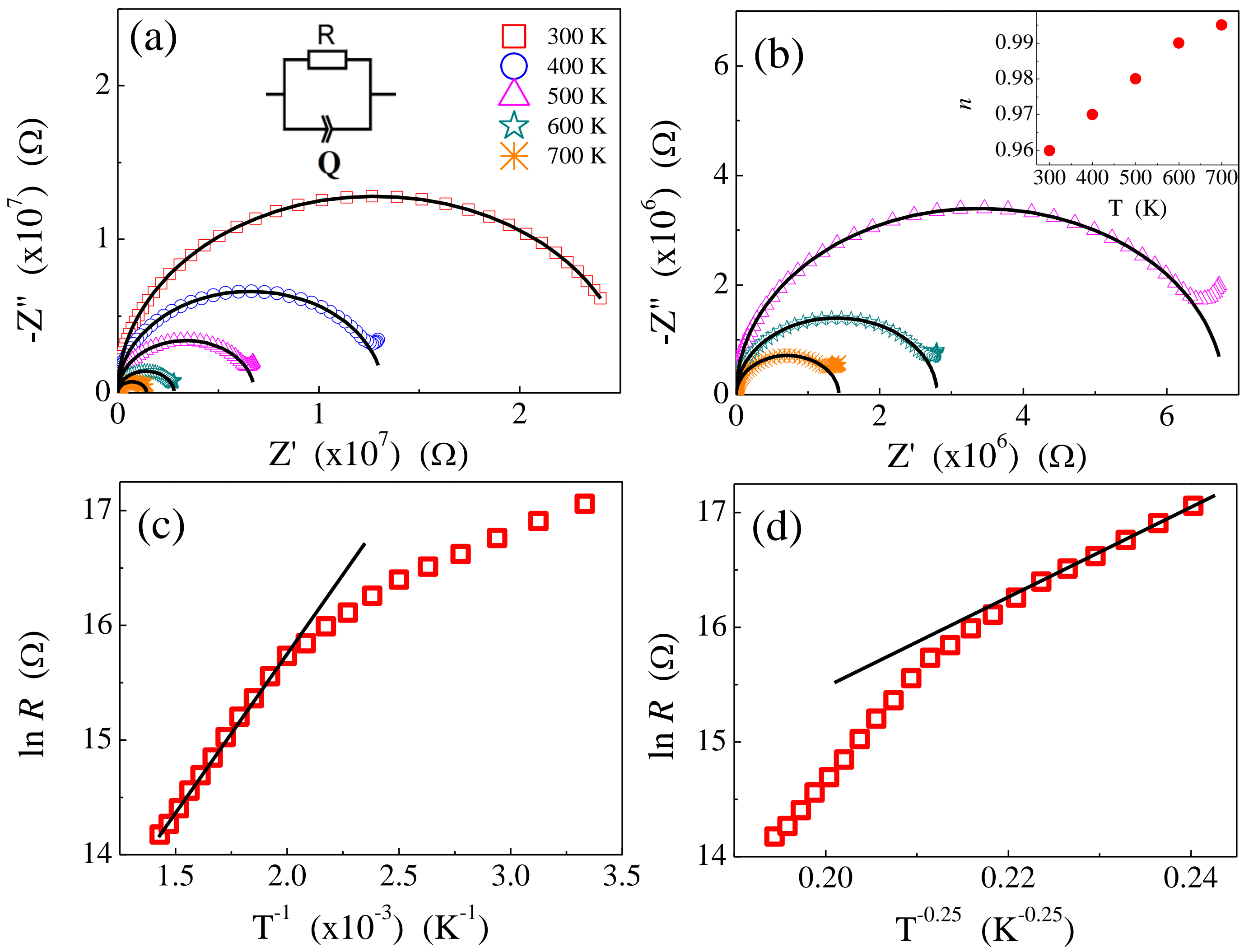

Fig. 6: Z-plots at different temperatures for PMT (a, b). Solid lines are fitted data to the observed data (points). Inset (a) shows the equivalent circuit used to fit the observed data. Inset (b) shows the temperature variation of *n*. Bulk resistance (*R*) is plotted using (c) NNH model and (d) MVRH model.

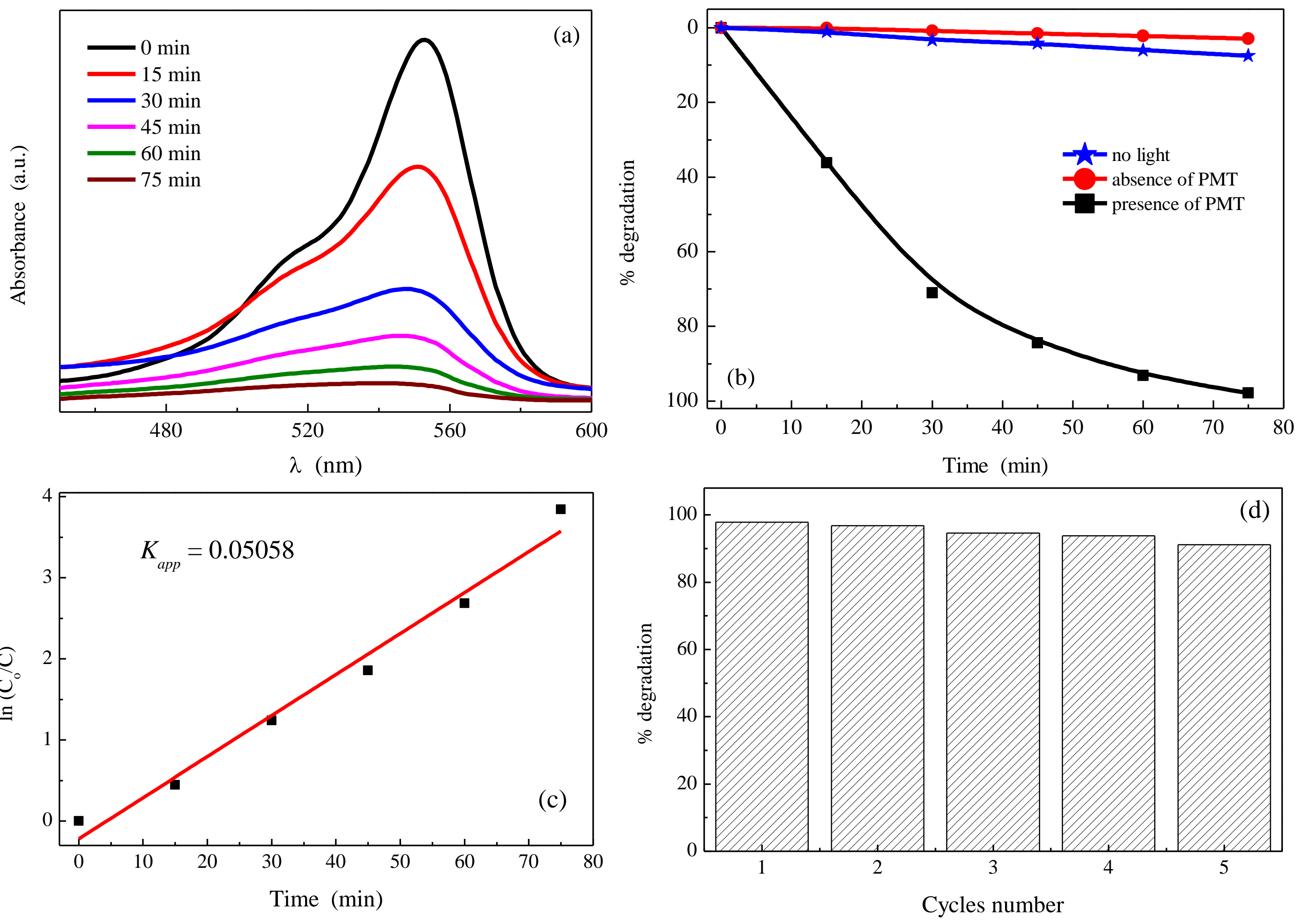

Fig. 7: (a) Change in absorbance of Rh-B solution after different irradiation time in presence of PMT. (b) % degradation of Rh-B in presence of UV only (red circles), PMT in the dark (blue star) and PMT under UV light (black square). (c) ln[$C_0$/C] as a function of irradiation time for PMT. (d) Cyclic runs in the %degradation of Rh-B using PMT under UV light.

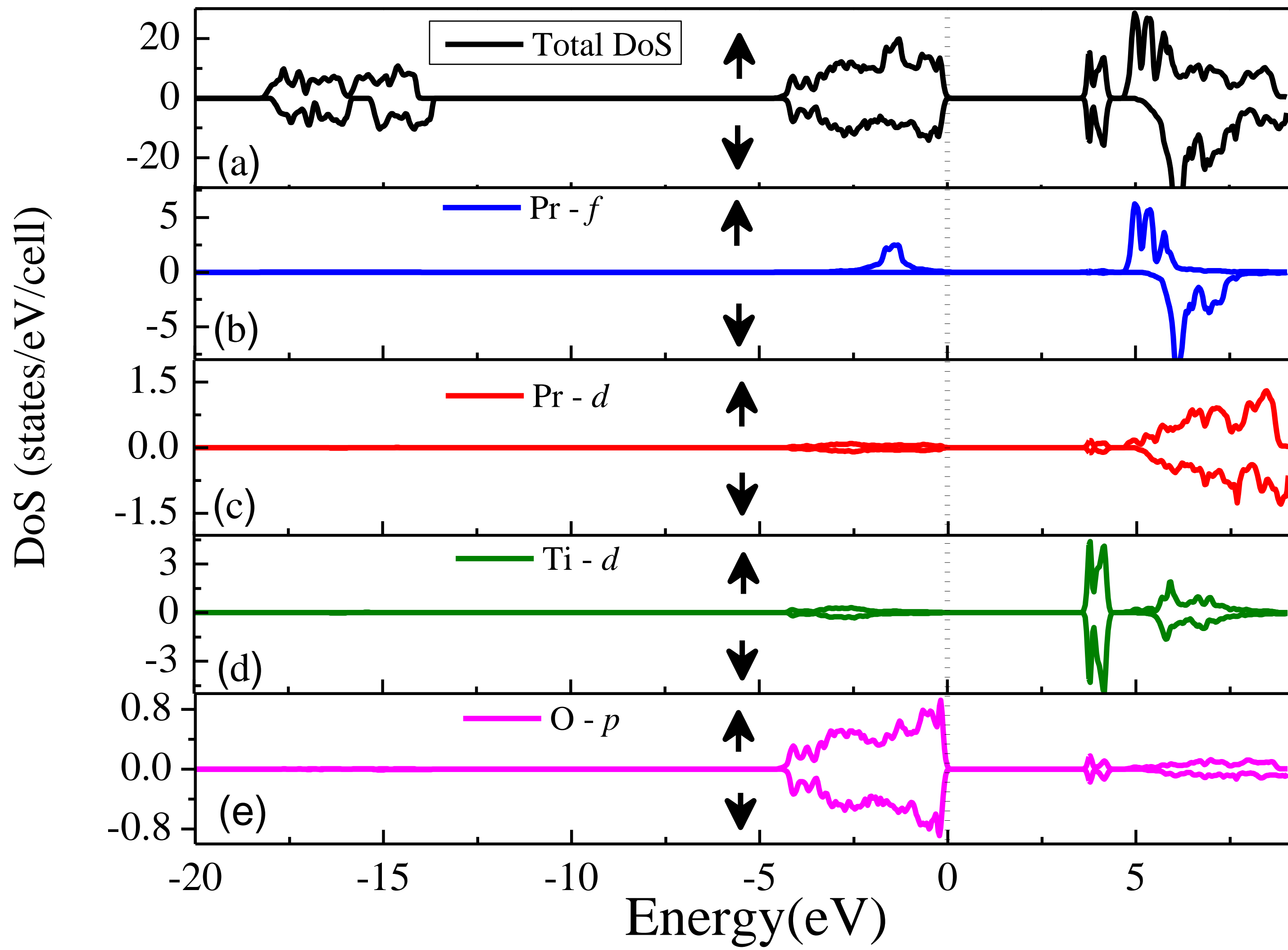

Fig. 8: Calculated total DoS (a) and partial DoS of Pr*f* (b), Pr *d* (c), Ti *d* (d) and O *p* (e) states of PMT.

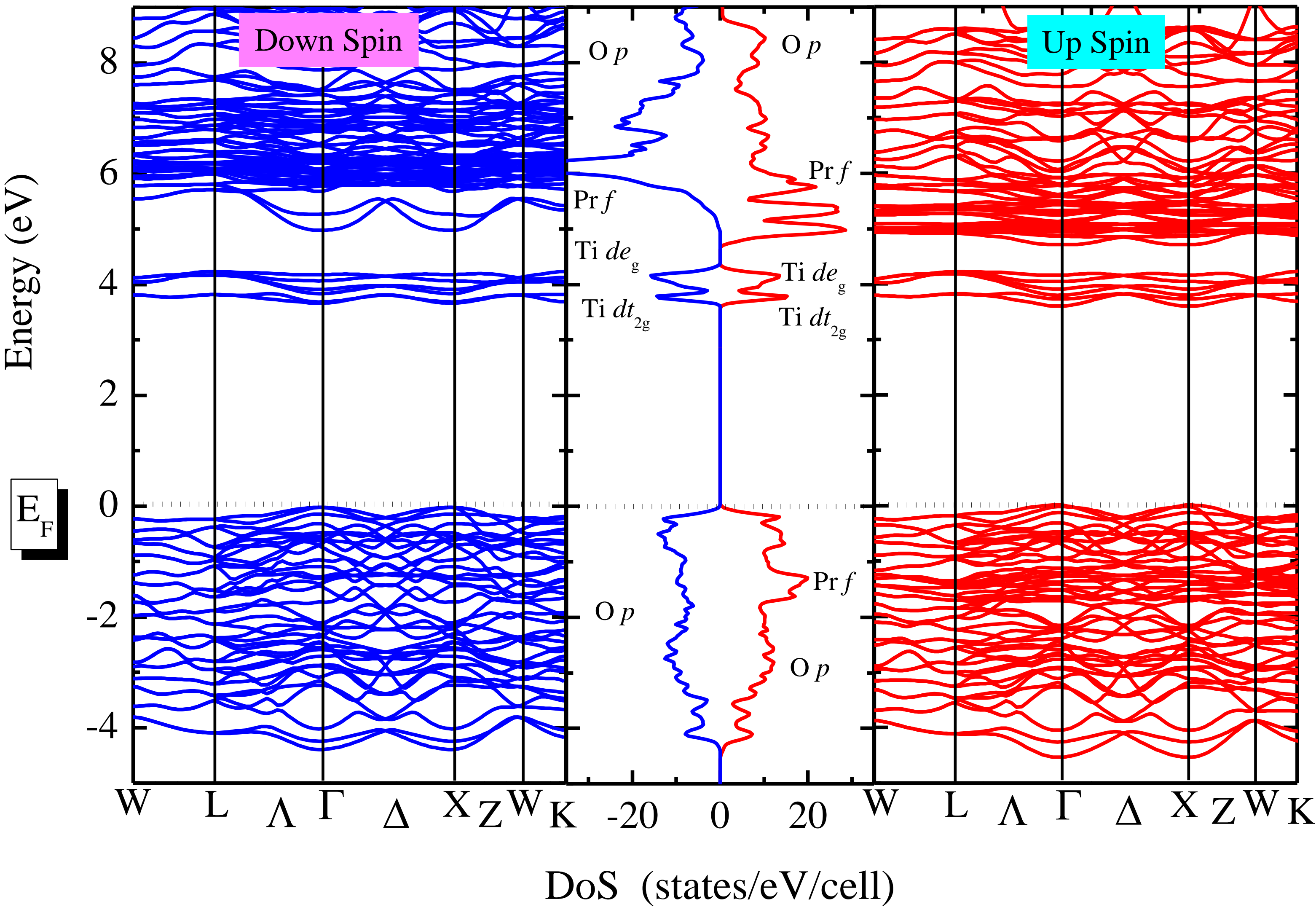

Fig. 9: Band structure for PMT using GGA+U. Bands in red and blue indicate spin up and spin down, respectively.

**Figure Captions:**

Fig. 1: (a) The X-ray diffraction pattern of PMT at RT, inset shows the unit cell of PMT. (b) The SEM micrograph of PMT.

Fig. 2: The photoluminescence excitation (a) and emission (b) spectra of PMT at RT. (c) The schematic energy level diagram of the PMT. (d) The UV-vis absorbance spectrum of PMT and the inset shows the Tauc plot (direct allowed transition) calculated from the UV-vis spectrum.

Fig. 3: The FTIR spectrum of PMT.

Fig. 4: (a) Frequency dependence of the ac conductivity at various temperatures. (b) The temperature variation of dc conductivity for PMT; The dc conductivity of PMT describes the conduction process using (c) NNH model, (d) MVRH model; The insets show the hopping energy (d) as a function of temperature; (c) Variation of activation energy with the temperature below 400 K.

Fig. 5: Frequency dependence of $\varepsilon'$ (a) and $\tan\delta$ (b) at various temperatures for PMT. Temperature dependence of $\varepsilon'$ (c) at various frequencies for PMT. Inset (c) shows the closed view of $\varepsilon'$ with temperature. Relaxation times $\tau$ of the charge carriers at bulks are plotted using (d) NNH model and the inset shows $\tau$ is plotted using (d) MVRH model.

Fig. 6: Z-plots at different temperatures for PMT (a, b). Solid lines are fitted data to the observed data (points). Inset (a) shows the equivalent circuit used to fit the observed data. Inset (b) shows the temperature variation of $n$. Bulk resistance ($R$) is plotted using (c) NNH model and (d) MVRH model.

Fig. 7: (a) Change in absorbance of Rh-B solution after different irradiation time in presence of PMT. (b) % degradation of Rh-B in presence of UV only (red circles), PMT in the dark (blue star) and PMT under UV light (black square). (c) $\ln[C_0/C]$ as a function of irradiation time for PMT. (d) Cyclic runs in the %degradation of Rh-B using PMT under UV light.

Fig. 8: Calculated total DoS (a) and partial DoS of Pr *f* (b), Pr *d* (c), Ti *d* (d) and O *p* (e) states of PMT.

Fig. 9: Band structure for PMT using GGA+U. Bands in red and blue indicate spin up and spin down, respectively.

Manuscript number: **PHYSB-D-20-02735**

Title: **Hopping mechanisms, photoluminescence studies toward highly efficient UV-responsive $Pr_2MgTiO_6$ photocatalyst.**

**Tables:**

Table 1: Structural parameters extracted from the Rietveld refinement of the XRD data for PMT.

**Space group:** *Pnma* (Orthorhombic)

**Cell parameters:** $a$ = 5.5710 Å, $b$ = 7.7893 Å, and $c$ = 5.4889 Å.

**Reliability factors:** $R_{exp}$ = 6.77, $R_p$ = 5.39, $R_{wp}$ = 7.06 and $\chi^2$= 1.09.

| **Atoms** | **Wyckoff site** | ***x*** | ***y*** | ***z*** | **Bond length (Å)** | **Bond angle (º)** |
|---|---|---|---|---|---|---|
| Pr | 4*c* | 0.54063 | 0.25000 | 0.51200 | Mg–O1 = 2.001(10) | Mg–O1–Ti = 153.3(4) |
| Mg | 4*b* | 0.50000 | 0.00000 | 0.00000 | Mg–O2 = 1.91(5) | Mg–O2–Ti = 156.2(18) |
| Ti | 4*b* | 0.50000 | 0.00000 | 0.00000 | Mg–O2 = 2.09(4) | |
| O1 | 4*c* | -0.02601 | 0.25000 | 0.41993 | Ti–O1 = 2.001(10) | |
| O2 | 8*d* | 0.29841 | 0.03811 | 0.72531 | Ti–O2 = 1.91(5) | |
| | | | | | Ti–O2 = 2.09(4) | |

Table 2: Fitted parameters are collected from ac conductivity spectra for PMT.

| **T(K)** | $\sigma_{dc}$ **(×10$^{-7}$) (S/m)** | ***A*(×10$^{-9}$)** | $k_1$ | ***B* (×10$^{-14}$)** | $k_2$ |
|---|---|---|---|---|---|
| 300 | 2.09676 | 0.78 | 1.2 | 11000 | 1.12 |
| 400 | 30.3991 | 3.9 | 0.99 | 0.18 | 1.79 |
| 500 | 156.261 | 4.2 | 0.99 | 0.13 | 1.78 |
| 600 | 633.51 | 5 | 0.99 | 0.14 | 1.75 |
| 700 | 1489.87 | 5 | 0.99 | 0.15 | 1.73 |

**Table captions:**

Table 1: Structural parameters extracted from the Rietveld refinement of the XRD data for PMT.

Table 2: Fitted parameters are collected from ac conductivity spectra for PMT.